\documentstyle[12pt]{article}

\def\sk{\vskip 1em}
\def\noi{\noindent}

\def\g{\gamma}
\def\th{\theta}

\def\Fh{\widehat{F}}
\def\CLh{\widehat{\CL}}

\def\del{\delta}
\def\D{\delta}

\def\sma#1{\mbox{\footnotesize #1}}

\def\11{\mbox{$1$}}
\def\pa{\partial}
\def\Cal{\cal}
\def\r{\rho}
\def\s{\sigma}

\def\det{{\rm{det}}}
\def\CL{{\cal{L}}}
\def\CF{{\cal{F}}}
\def\CK{{\cal{K}}}
\def\ap{{\alpha^\prime}}
\def\DBI{{\rm{BI}}}
\def\BI{{\rm{BI}}}
\def\D3{{\rm{D3}}}

\def\2{{\geq 2}}
\def\*{\star}

\renewcommand{\thefootnote}{\alph{footnote}}

\newcommand{\eq}{\begin{equation}}
\newcommand{\en}{\end{equation}}
\newcommand{\beeq}{\begin{equation}}
\newcommand{\beqn}{\begin{equation}}
\newcommand{\eeqn}{\end{equation}}
\newcommand{\eneq}{\end{equation}}
\newcommand{\beqr}{\begin{eqnarray}}
\newcommand{\eeqr}{\end{eqnarray}}

\newcommand{\matc}{\begin{array}{c}}
\newcommand{\matcc}{\begin{array}{cc}}
\newcommand{\matccc}{\begin{array}{ccc}}
\newcommand{\matcccc}{\begin{array}{cccc}}
\newcommand{\emat}{\end{array}}

\newcommand{\IH}{\relax{\rm I\kern-.18em H}}
\newcommand{\IR}{\relax{\rm I\kern-.18em R}}
\newcommand{\IK}{\relax{\rm I\kern-.18em K}}
\newcommand{\II}{\hbox{\rm 1\kern-.35em 1}}
\newcommand{\Is}{\relax{\rm 1\kern-.35em 1}}

\begin{document}


\vskip -0.55cm 
\hfill  ${}^{{}^{{}^{\displaystyle{\mbox{LMU-TPW 01-02}}}}}$   

\vskip 0.95cm
\renewcommand{\thefootnote}{\fnsymbol{footnote}}
\vskip 0.85cm
\noi{\Large\bf  On Duality Rotations 
in Light-Like\\ Noncommutative
Electromagnetism}\footnote{
Talk presented at the 
Euroconference: Brane New World and Noncommutative Geometry, 
Villa Gualino (Torino), October 2000}
\vskip 0.40cm
\noi{\large Paolo Aschieri}\\
{\it Sektion Physik der Ludwig-Maximilians-Universit\"at\\
Theresienstr. 37, D-80333 M\"unchen, Germany\\ 
\small{\it{e-mail address: aschieri@theorie.physik.uni-muenchen.de}}}
\vskip 0.40cm
\noi {ABSTRACT:}
We study electric-magnetic duality rotations for 
noncommutative electromagnetism (NCEM).
We express NCEM as a nonlinear commutative $U(1)$ gauge theory
and show that it is self-dual when the noncommutativity parameter $\th$
is light-like (e.g. $\th^{0i}=\th^{1i}$). This implies, in the slowly 
varying field approximation, self-duality of NCEM to all orders in 
$\th$.
\vskip 0.15cm
\noi{\small Keywords: Duality; 
Born-Infeld; Noncommutative Gauge Theory }

\vskip 0.40cm


\noi{\bf{Introduction}}\\
Field theories on noncommutative spaces have received 
renewed interest since their relevance in describing 
Dp-branes effective actions (see \cite{Seiberg:1999vs} and 
references therein).  
Noncommutativity in this context is due to a nonvanishing 
NS background two form on the Dp-brane. Initially 
space-like (magnetic) backgrounds ($B^{ij}\not= 0$) 
were considered, 
then NCYM theories also with time 
noncommutativity ($B^{0i}\not=0$) have been studied 
\cite{Seiberg:2000gc}.
It turns out that unitarity of NCYM holds only if $B$ is 
space-like or light-like (e.g. $B_{0i}=-B_{1i}$) and that these are
precisely the NCYM theories that can be obtained from open strings in the 
decoupling limit $\ap\rightarrow 0$ \cite{Aharony:2000gz}.
Following \cite{Seiberg:1999vs}, gauge 
theory on a Dp-brane with constant two-form $B$  
can be described via a commutative Lagrangian and field strength
$\CL(F+B)$ or via a  noncommutative one  $\widehat{\CL}(\Fh)$,
where $\Fh_{\mu\nu}=\pa_\mu A_\nu - \pa_\nu A_\mu 
-i[ A_\mu\*_{\!\!\!{\textstyle{,}}} A_\nu]$. (Here $\*$ is the star 
product,  on coordinates
$[x^\mu\*_{\!\!\!{\textstyle{,}}}x^\nu]=
x^\mu\*x^\nu-x^\nu\*x^\nu=i\th^{\mu\nu}$, where $\th$ depends 
on $B$  and the metric on the Dp-brane).          
These two descriptions are complementary and are related by
Seiberg-Witten map (SW map). In the $\ap\rightarrow 0$ 
limit \cite{Seiberg:1999vs} the exact 
effective electromagnetic theory on a Dp-brane is
NCEM, this is equivalent, via SW map, 
to a nonlinear commutative $U(1)$ gauge 
theory. For a D3-brane, in the slowly varying field approximation, 
we give an explicit expression of this nonlinear $U(1)$ theory and 
we show that it is self-dual when $B$ (or $\th$) is light-like. 
Via SW map solutions of $U(1)$ nonlinear 
electromagnetism are mapped into solutions of NCEM, so that
duality rotations are  also a symmetry of NCEM, i.e. NCEM is self-dual.
When $\th$ is space-like we do not have self-duality and the 
S-dual of space-like NCYM is a 
noncommutative open string theory decoupled from closed strings 
\cite{Gopakumar:2000na}. Related work appeared 
in \cite{Rey:2001hh}.
Self-duality  of NCEM was initially studied in
\cite{Ganor:2000my} to first order in $\th$. 

We show self-duality of NCEM using Gaillard-Zumino approach 
\cite{Gaillard:1981rj} to study duality 
rotations in nonlinear electromagnetism, we thus provide 
[see (\ref{L})] a 
new example of  Lagrangian satisfying Gaillard-Zumino self-duality 
condition. We present here
the case where the axion and the dilaton are zero. It is also 
possible to include arbitrary constant axion and dilaton
as well as a kinetic and interaction term for Higgs fields. 
Higgs fields in the noncommutative description result minimally coupled 
to the gauge field.
Formally (covariant derivatives, minimal couplings) NCEM
resembles commutative $U(N)$ YM, and on tori with rational $\th$
the two theories are $T$-dual \cite{Schwarz:1998qj}.
Self-duality of NCEM then hints to a possible duality symmetry 
of the equations of motion of $U(N)$ YM.

This paper is organized as follows. We first review Gaillard-Zumino 
self-duality condition and see that the D3-brane Lagrangian is
self-dual \cite{Tseytlin:1996it}. We then present a simple argument 
showing why we need $B$ and $\th$ light-like in the zero slope limit.
Finally we discuss self-duality of noncommutative Born-Infeld theory
and NCEM.
\sk
\noi{\bf{Duality Rotations}}

\noi Consider in four dimension and with metric $g_E$ the
Lagrangian density (Lagrangian for short) 
\eq
{\Cal{L}}(F_{\mu\nu}, g_{E\,\mu\nu}, \chi^i)
=\sqrt{-g_E}\,
L(F_{\mu\nu}, g_{E\,\mu\nu}, \chi^i)
\en
where $\chi^i$ are some constant 
parameters that can possibly have space-time 
indices. If we define\footnote{The Hodge dual of a two form 
$\Omega_{\mu\nu}$ is defined by 
$\Omega^*_{\mu\nu}\equiv \frac{1}{2}\sqrt{-g_E}\,
\epsilon_{\mu\nu\rho\sigma}
\Omega^{\rho\sigma}$, where $\epsilon^{0123}=-\epsilon_{0123}=1$;
$g_E$ has signature $(-,+,+,+)$. 
We have 
${\Omega^*}^*_{\mu\nu}=-\Omega_{\mu\nu}$,
$\Omega^{*\,\mu\nu}= \frac{1}{2}\sqrt{-g_E^{-1}}\, 
\epsilon^{\mu\nu\rho\sigma}\Omega_{\rho\sigma}$.}
 
\beeq
 {K^*}^{\mu\nu}~=~\frac{\pa L}{\pa F_{\mu\nu}}~~~~~~~~~~~~~~
{\left({ {\partial F_{\rho\sigma}}\over{\partial F_{\mu\nu}}  } 
=\delta^{\mu}_{\rho}\delta^{\nu}_{\sigma}-
\delta^{\nu}_{\rho}\delta^{\mu}_{\sigma}\right)}~  
\label{defK}
\eneq
then the Bianchi identity and the equations of motion (EOM) for $\CL$ 
read
\beqr
\begin{array}{c} 
\pa_{\mu}(\sqrt{-g_E}\, F^{*\,\mu\nu} )=
{\pa}_{\mu}
\tilde{F}^{\mu\nu}  =0~,\\[0.8em]
{\pa}_{\mu}(\sqrt{-g_E}\,  K^{*\,\mu\nu} )=
{\pa}_{\mu}
\tilde{K}^{\mu\nu}  =0~,
\end{array}\label{3}
\eeqr
where ${\tilde{F}}^{\mu\nu}
\equiv 
\frac{1}{2}\epsilon^{\mu\nu\rho\sigma} F_{\rho\sigma}$ and 
${\tilde{K}}^{\mu\nu}
\equiv 
\frac{1}{2}\epsilon^{\mu\nu\rho\sigma} K_{\rho\sigma}$.
Under the
infinitesimal transformations 
\beeq
\delta
\left(
\matc
K \\
F
\emat
\right)
=
\left(
\matcc
\mbox{A} & \mbox{B} \\
\mbox{C} & \mbox{D}
\emat
\right)
\left(
\matc
K \\
F
\emat
\right)~\label{FKtrans}
\eeqn
the system of EOM (\ref{3}) is mapped into itself. 
We can also allow the parameters $\chi$ to vary
\beeq
\delta \chi^i=\xi^i(\chi)~,
\label{transchi}
\eeqn
then under (\ref{FKtrans}),(\ref{transchi}) 
the $\CL_\chi$ EOM (\ref{3}) are 
mapped into the
$\CL_{\chi+\del\chi}$ EOM. 
Consistency of (\ref{FKtrans}),(\ref{transchi}) 
with the definition of $K$, i.e.
$K+\del K=\frac{\pa}{\pa(F+\del F)}\CL(F+\del F,g,\chi+\del\chi)$ 
holds in particular if 
$~\!
\left( {}^{\mbox{\sma {A}} 
\mbox{\sma {B}}}_{\mbox{\sma {C}} 
\mbox{\sma{D}}} \right)$ 
belongs to the Lie algebra of $SL(2,R)$,
and the variation of the Lagrangian under 
(\ref{FKtrans}),(\ref{transchi})
can be written as
\eq
\delta {\Cal{L}}\equiv(\delta{_F}+\delta{_\chi}){\Cal{L}}
={1\over 4}(\mbox{B}F\tilde{F}+\mbox{C}K\tilde{K})
 \label{cond}~.
\en
When (\ref{cond}) holds,  under a finite $SL(2,R)$ rotation
a solution $F$ of $\CL_\chi$ is mapped
into a solution $F'$ of $\CL_{\chi'}$ and we say that $\CL_\chi$
is self-dual. A self-duality condition equivalent to (\ref{cond}) 
is obtained using 
(\ref{FKtrans}) to evaluate the $\del_F\CL$ term in (\ref{cond})
\eq
\del_\chi\CL=\frac{1}{4}{\rm{B}}F\tilde{F}
           -\frac{1}{4}{\rm{C}}K\tilde{K}
           -\frac{1}{2}{\rm{D}}{F}\tilde{K}
\label{cond1}~~.
\en
If the $\chi$ parameters are held fixed (strict self-duality)
the maximal duality group is $U(1)$ 
\cite{Gaillard:1981rj}, see also the nice review \cite{Kuzenko:2001uh}.
Viceversa we can always extend a $U(1)$ self-dual Lagrangian to a 
$SL(2,R)$ self-dual one introducing two real valued scalars 
$S=S_1+iS_2$ (axion and dilaton).
If the Lagrangian $\CL(F)$ is self dual under $U(1)$, then the new Lagrangian
\eq
\check{\CL}(F,S)\equiv\CL(S_2^{\frac{1}{2}}F)+\frac{1}{4}S_1F\tilde{F}
\label{noncomp}~,
\en
where $F\tilde{F}\equiv F_{\mu\nu}\tilde{F}^{\mu\nu}$, is self-dual under 
$SL(2,R)$ provided that $S=S_1+iS_2$ transforms as 
\eq
S'=\frac{aS+b}{cS+d}~~~~~~{\rm{where}} ~~~
\left( {}^a_c{~}^b_d \right)\in SL(2,R)~.
\en
The term in (\ref{noncomp}) proportional to $S_1$ is a total derivative if $S_1$ is 
constant; this term does not affect the EOM but enters the definition 
of $K$. To recover the original Lagrangian just set $S_1=0$ and $S_2=1$;
the duality group is then $U(1)$, the $SL(2,R)$ subgroup that leaves
$S_1=0$ and $S_2=1$ invariant.

\sk
We now discuss self-duality of 
the D3-brane effective action in a IIB supergravity background 
with constant axion, dilaton NS and RR two-forms. 
The background two-forms can be gauged away in the bulk and we are left
with the field strength $\CF=F+B$ on the D3-brane. 
Here $B$ is defined as the constant part of $\CF$, or
$B=\CF|_{\rm{spatial}\, \infty}$ since 
$F$ vanish at spatial infinity.
{}For slowly varying fields
the Lagrangian, 
in string and in Einstein frames, respectively reads\footnote{we omit the RR four-form $C_4$ because it is invariant under $SL(2,R)$ 
duality rotations. 
If $\CL$ is self-dual then also $\CL_{\rm{D3}}=\CL+\tilde{C}_4$ 
is self-dual (here 
$\tilde{C}_4=\frac{1}{24}\epsilon^{\mu\nu\r\s}C_{\mu\nu\r\s}$).}
\beqr
\CL&=&
\frac{-1}{\ap^2 g_s}\sqrt{-{\rm{det}}(g+\ap {\CF})}
+\frac{1}{4}C{\CF}\tilde{\CF}\nonumber\\[1.2em]
&=&
\frac{-1}{\ap^2}\sqrt{-{\rm{det}}(g_E+\ap S_2^{1/2}{\CF})}
+\frac{1}{4}S_1{\CF}\tilde{\CF}\label{LDBIWZ}\\[1.2em]
&=&\frac{-1}{\ap^2}\sqrt{-g_E}
\; {\sqrt{1+\frac{\ap^2}{2} S_2 {\CF}^{\,2}-
\frac{\ap^4}{16}S_2^2(\CF\CF^*)^2}}+
\frac{1}{4}S_1{\CF}\tilde{\CF}\nonumber
\eeqr
in the second line
$S=S_1+iS_2=C+\frac{i}{g_s} \;$
while, in the last line, we have simply expanded the 4x4 determinant 
and 
$\CF^{\,2}\equiv\CF_{\mu\nu}\CF^{\mu\nu}\,,~ \CF\CF^*\equiv\CF_{\mu\nu}
\CF^{*\,\mu\nu}$.

Under the $SL(2,R)$ rotation
\beeq
\left(
\matc
\CK^\prime \\
\CF^\prime
\emat
\right)
=
\left(
\matcc
a & b \\
c & d
\emat
\right)
\left(
\matc
\CK \\
\CF
\emat
\right)~,~~S^\prime=\frac{aS+b}{cS+d}\label{Srot}~,~~g_E'=g_E~,~~(\ap)'=\ap~,
\label{F+Brot}
\eeqn
where $\tilde{\CK}=\frac{\pa}{\pa \CF}\CL$, it is not difficult to directly 
check that the 
Lagrangian  $\CL$ satisfies the self-duality condition (\ref{cond})
(with $F,K$ replaced by $\CF,\CK$).
{}For  simplicity, 
we will later set $S_1=0$ and $S_2=g_s=1$. Then the duality group 
reduces to $U(1)$,
moreover string and Einstein frames coincide.  
The Lagrangian reduces to the Born-Infeld Lagrangian
\eq
\CL_{\rm{BI}}=
\frac{-1}{\ap^2}\sqrt{-{\rm{det}}(g+\ap {\CF})}
\label{BI}~.
\en
Using (\ref{noncomp}) we can always recover the more 
general situation (\ref{LDBIWZ}).
The  explicit expression of $\CK$ is 
\eq
{\CK}_{\mu\nu}=
\frac{\CF^*_{\,\mu\nu}+\frac{\ap^2}{4}\CF\CF^*\,\CF_{\mu\nu}}{
\sqrt{1+\frac{\ap^2}{2} \CF^2-\frac{\ap^4}{16}(\CF\CF^*)^2}}~.
\label{Ktransf}
\en
From (\ref{F+Brot}) and (\ref{Ktransf}), one can extract how $B$ 
(the constant part of $\CF$) transforms
\eq
{B'}_{\mu\nu}=\cos\!\gamma\, B_{\mu\nu} - \sin\!\gamma\,
\frac{B^*_{\,\mu\nu}+\frac{\ap^2}{4}B B^*\, B_{\mu\nu}}{
\sqrt{1+\frac{\ap^2}{2} B^2-\frac{\ap^4}{16}(B B^*)^2}}\label{Brot}~;
\en
this transformation is independent from the slowly varying fields
approximation.
\sk 

\noi{\bf{Open/closed strings and light-like noncommutativity}}
\label{Seiberg-Witten}

\noi The open and closed string parameters are related by 
(see \cite{Seiberg:1999vs}, 
the expressions for $G$ and $\th$ first appeared in 
\cite{Chu:1999qz}) 
\eq
\begin{array}{l}
\displaystyle
\frac{1}{g+\ap B}=G^{-1}+\frac{\th}{\ap} ~~\nonumber\\[1em]
\displaystyle
g^{-1}=(G^{-1}-\th/\ap)\,G\,(G^{-1}+\th/\ap)=G^{-1}-\ap^{-2}
\th\, G\, \th\nonumber\\[1em]
\displaystyle
\ap B=-(G^{-1}-\th/\ap)\,\th/\ap\,(G^{-1}+\th/\ap)\nonumber\\[1em]
\displaystyle
G_s=g_s 
\sqrt{\frac{{\rm{det}}G}{\det(g+\ap B)}}=g_s\sqrt{\det G\;
\det{(G^{-1}+\th/\ap)}}=g_s\sqrt{\det g^{-1}\;
\det{(g+\ap B)}}
\end{array}
\label{Gsgs}
\en 
The decoupling limit 
$\ap\rightarrow 0$ with $G_s,G,\th$ 
nonzero and finite 
\cite{Seiberg:1999vs}  
leads to a well defined field theory only if
$B$ is space-like or
light-like \cite{Aharony:2000gz}. 
Looking at the closed and open string coupling constants it is 
easy to see why one needs this space-like or light-like condition on 
$B$.
Consider the coupling constants ratio $G_s/g_s$, that
expanding the 4x4 determinant reads 
(here $B^2=B_{\mu\nu}B_{\r\s}g^{\mu\r}g^{\nu\s}$, 
$\th^2=\th^{\mu\nu}\th^{\r\s}G_{\mu\r}G_{\nu\s}$ and so on)
\eq
\frac{G_s}{g_s}=\sqrt{1+\frac{\ap^{-2}}{2} 
\th^2-\frac{\ap^{-4}}{16}(\th \th^*)^2} 
\;=\;
\sqrt{1+\frac{\ap^2}{2} B^2-\frac{\ap^4}{16}(B B^*)^2}\label{Gs/gs}~.
\en
Both $G_s$ and $g_s$ must be positive; since $G$ and $\th$ are by
definition 
finite for $\ap\rightarrow 0$ this implies 
$\th \th^*=0$ and $\th^2\geq 0$. Now  $\th \th^*=0 \Leftrightarrow 
\det\th=0 \Leftrightarrow \det B=0 \Leftrightarrow B B^*=0$.
In this case from (\ref{Gs/gs}) we also have $\th^2=\ap^4B^2$.
In  conclusion the $\ap\rightarrow 0$ limit defined by
keeping $G_s,G,\th$ nonzero and finite \cite{Seiberg:1999vs}, 
is well defined iff
\eq 
B^2\geq 0~,~~B B^*= 0
~~~~~~~\mbox{i.e.}~~~~~~~\th^2\geq 0~,~~\th \th^*=0
\label{Bth}
\en
This is the condition for $B$ (and 
$\th$) to be space-like or light-like.
Indeed with Minkowski metric (\ref{Bth}) reads 
$\vec{B}^2-\vec{E}^2\geq 0$ and $\vec{E}\perp\vec{B}$. 

If we now require the $\ap\rightarrow 0$ limit to be compatible with 
duality rotations, we immediately see that we have to consider only the 
light-like case $B^2=  B B^*= 0$. Indeed under $U(1)$ rotations 
the electric and magnetic fields mix up, in particular under a $\pi/2$
rotation (\ref{Brot}) a space-like $B$ becomes time-like. 

In the light-like case,  relations (\ref{Gsgs}) simplify considerably.
The open and closed string coupling constants coincide: 
$G_s=g_s=S_2^{-1}=1$.
Use of the relations
\eq
\Omega^*_{\mu\r}{\Omega^*}^{\r\nu}-
\Omega_{\mu\r}\Omega^{\r\nu}=
\frac{1}{2}\Omega^2\,\del_{\mu}^{~\nu}~\,,\,~~ 
\Omega_{\mu\r}{\Omega^*}^{\r\nu}=
{\Omega^*}_{\mu\r}\Omega^{\r\nu}=
\frac{-1}{4}\Omega\Omega^*\,\del_\mu^{~\nu}\label{S}
\en
valid for any antisymmetric tensor $\Omega$, shows that 
any two-tensor at least cubic in $\th$ (or $B$) vanishes.
It follows that $g^{-1}G\,\th=\th$ and that the 
raising or lowering of the  
$\th$ and $B$ indices is independent from the metric used.
We also have
\eq
B_{\mu\nu}=-{\ap}^{-2}\th_{\mu\nu}\label{Btheta}~~.~~ ~
\en
\sk
\noi{\bf{Self-duality of NCBI and NCEM}}

\noi We now study duality rotations for noncommutative 
Born-Infeld (NCBI) theory and its zero slope limit that is NCEM.
The relation between the NCBI and the BI Lagrangians is 
\cite{Seiberg:1999vs}
\eq
\CLh_\BI(\Fh,G,\th,G_s)=\CL_\BI(F+B,g)+O(\pa F)+\rm{tot.der.}
\label{NCDBI}
\en
where $O(\pa F)$ stands for higher order derivative corrections, 
$\Fh$ is the noncommutative $U(1)$ field strength and we have set 
$g_s=1$.
The NCBI Lagrangian is
\eq
\CLh_\DBI(\Fh,G,\th,G_s)=
\frac{-1}{\ap^2 G_s}\sqrt{-{\rm{det}}(G+\ap {\Fh})}
+O(\pa\Fh)~.
\en
In the slowly varying field approximation the action of duality 
rotations on $\CLh_\BI$  
is derived from self-duality of $\CL_\BI$. 
If $\Fh$ is a solution of the 
$\CLh_\DBI^{G_s,G,\th}$ EOM then $\Fh'$ obtained via 
$\Fh\stackrel{\sma{\rm{SW map}}}{\longleftarrow\!\!\!\longrightarrow}
\CF\stackrel{\sma{\rm{duality\, rot.}}}{\longleftarrow\!\!\longrightarrow}
\CF'
\stackrel{\sma{\rm{SW map}}}{\longleftarrow\!\!\!\longrightarrow}
\Fh'$ is a solution of the 
$\CLh_\DBI^{G^\prime_{\!s},G',\th'}$ EOM where $G^\prime_s,G',\th'$
are obtained using (\ref{Gsgs}) from $g'$,  $B'$ and $g_s'=g_s=1$. 

In the light-like case we have $G_s=g_s=1$, the
$B$ rotation (\ref{Brot}) simplifies to
\eq
B^\prime_{\mu\nu}=\cos\!\gamma\,B_{\mu\nu}-\sin\!\gamma\,
B^*_{\,\mu\nu}~,\label{Bsrot}
\en
using (\ref{Srot}) and (\ref{Bsrot}) the $U(1)$ duality  
action on the open  string variables is 
\eq
G^\prime=G~~,~~~\th^{\prime\,\mu\nu}=
\cos\!\g\,\th^{\mu\nu}-\sin\!\g\,{\th^*}^{\mu\nu}~.
\label{Tsrot}
\en
For $\th$ light-like, solutions $\Fh$ of $\widehat{\CL}^{G,\th}$ 
are mapped
into solutions $\Fh'$ of $\widehat{\CL}^{G,\th'}$ and therefore 
$\widehat{\CL}^{G,\th}$ is self-dual.
Moreover, by a rotation in three dimensional space we can 
map $\th'$ into $\th$.
In order to show self-duality of NCEM we consider the
zero slope limit of (\ref{NCDBI}) and verify that  
the resulting lagrangian
on the r.h.s. of (\ref{NCDBI}) is self-dual.
We
rewrite $\CL_\BI$ in terms of the open string parameters $G,\th$
\beqr
\CL_\BI&=&\frac{-1}{\ap^2 }\sqrt{-{\rm{det}}(g+\ap\CF)}
=\frac{-\sqrt{G}}{\ap^2 }\sqrt{\frac{{\rm{det}}(g+\ap B+\ap{F})}
{{\rm{det}}(g+\ap B)}}\nonumber\\[1.2em]
&=&\frac{-1}{\ap^2 }\sqrt{-{\rm{det}}(G+\ap{F}+G\th F)}
\label{LGth}~.
\eeqr
The determinant in the last line can be evaluated as sum of
products of traces (Newton-Leverrier formula). Each trace can 
then be rewritten in terms of the six basic Lorentz invariants
$F^2,~F F^*,~F\th,~F\th^*,~\th^2=\th\th^*=0$, explicitly
$$
\begin{array}{l}
\rm{det}{G^{-1}}\,\rm{det}(G+\ap{F}+G\th F)
=(1-\frac{1}{2}\th F)^2+\ap^2[\frac{1}{2}F^2+\frac{1}{4}
\th F^*\;FF^*]-\ap^4(\frac{1}{4}FF^*)^2
\nonumber
\end{array}~\,
$$
Finally we take the $\ap\rightarrow 0$ limit of 
(\ref{LGth}), drop the infinite constant and  total derivatives  
and denote by $\CL_\th^{\ap\!\!\rightarrow 0}$ the resulting Lagrangian
\eq
\CL_\th^{\ap\!\!\rightarrow 0}=
\frac{-\frac{1}{4}F^2-
\frac{1}{8}\th F^*\,FF^*}{1-
\frac{1}{2}\th F}\label{L}~\,.
\en
We thus have an expression for NCEM in terms of 
$G,\th$ and $F$  
\eq
\widehat{\CL}_{\rm{EM}}
\equiv-\frac{1}{4}{\Fh}^2=\CL_\th^{\ap\!\!\rightarrow 0}+O(\pa F)+\rm{tot. ~der.}
\en
The Lagrangian (\ref{L})
satisfies the self-duality condition (\ref{cond1}) with $\chi=\th$ 
and A$=$D$=0$, C$=-$B and therefore NCEM is self-dual 
under the $U(1)$ duality
rotations  (\ref{Tsrot}) and $F'=\cos\!\g\,F-\sin\!\g\,K$.


\sk
\sk
\noi{\bf{Acknowledgements}}\\
I am indebted with Sergei M. Kuzenko and Stefan Theisen 
for many enlightening discussions.
I wish to thank  Branislav Jur\u{c}o, John Madore, Peter Schupp
and Harold Steinacker for fruitful discussions, 
and the organizers of the conference 
for the stimulating atmosphere.
 This work has been supported by 
Alexander von Humboldt-Stiftung. 

\end{document}